\documentclass[preprint,aps,showpacs,unsortedaddress]{revtex4}
\usepackage{epsfig}
\begin{document}
\title{Option pricing under stochastic volatility: the exponential Ornstein-Uhlenbeck model}
\author{Josep Perell\'o}
\email{josep.perello@ub.edu}
\affiliation{Departament de F\'{\i}sica Fonamental, Universitat de Barcelona,\\
Diagonal, 647, E-08028 Barcelona, Spain}
\author{Ronnie Sircar}
\email{sircar@princeton.edu}
\affiliation{Department of Operations Research and Financial Engineering, Princeton University, E-Quad, Princeton, New Jersey 08544}
\author{Jaume Masoliver}
\email{jaume.masoliver@ub.edu}
\affiliation{Departament de F\'{\i}sica Fonamental, Universitat de Barcelona,\\
Diagonal, 647, E-08028 Barcelona, Spain}
\date{\today}

\begin{abstract}
We study the pricing problem for a European call option when the volatility of the underlying asset is random and follows the exponential Ornstein-Uhlenbeck model. The random diffusion model proposed is a two-dimensional market process that takes a log-Brownian motion to describe price dynamics and an Ornstein-Uhlenbeck subordinated process describing the randomness of the log-volatility. We derive an approximate option price that is valid when (i) the fluctuations of the volatility are larger than its normal level, (ii) the volatility presents a slow driving force toward its normal level and, finally, (iii) the market price of risk is a linear function of the log-volatility. We study the resulting European call price and its implied volatility for a range of parameters consistent with daily Dow Jones Index data.
\end{abstract}
\pacs{89.65.Gh, 02.50.Ey, 05.40.Jc, 05.45.Tp}
\maketitle

\section{Introduction}

The picture that asset prices follow a simple diffusion process was first proposed by Bachelier in 1900 and his main motivation was precisely to provide a fair price for the European call option. The buyer of the call option has the right, but not the obligation, to buy the underlying asset from the seller of the option at a certain time (the expiration date) for a certain prestablished price (the strike price)~\cite{hull}.

After the introduction in 1959 of the geometric Brownian motion as a more refined market model for which prices cannot be negative, the volatility can be viewed as the diffusion coefficient of this random walk~\cite{osborne}. This simple assumption -- the volatility being a constant -- lies at the heart of Black-Scholes (BS) option pricing method~\cite{black.scholes}. Within the BS theory, the most direct technique constructs an equivalent martingale measure for the underlying asset process. This change of the probability measure guarantees a fair price for the option. The price of the European call is readily obtained: it simply consists of an average over the final payoff evaluated under the martingale measure, and appropriate discounting.

However, and especially after the 1987 crash, the geometric Brownian motion model and the BS formula were unable to reproduce the option price data of real markets. Several studies have collected empirical option prices in order to derive their implied volatilities (the volatility that agrees with the BS formula). These tests conclude that the implied volatility is not constant and seems to be well-fitted to a U-shaped function of the ratio between the price of the underlying asset and the strike price of the option (the so-called ``moneyness''). This phenomenon, known as the smile or smirk effect, the latter term due to the fact that the U-shape is usually asymmetric, shows the inadequacy of the geometric Brownian model. Nevertheless, the continuous-time framework provides several alternative models specially designed to explain, at least qualitatively, this effect. Among them, we highlight on the Stochastic Volatility (SV) models. These are two-dimensional diffusion processes in which one dimension describes the asset price dynamics and the second one governs the volatility evolution (see e.g. Refs.~\cite{hull87,Stein,Heston,scott,fouque,fouquebook,bouchaud,ijtaf,qf}).

The SV class of models have also been studied to check whether they are also able to describe the dynamics of the asset itself. In this direction there are several interesting works in the literature focusing on this and others similar issues (see e.g. Refs.~\cite{spagnolo,bouchaud-lev,bacry,qiu,remer,pre-lev,comp,extreme,eisler,buchbinder,biro,montagna,pere2007}). It is indeed worth noticing that the theoretical framework of SV modeling has many common points with the research on random diffusion whose aim is to describe the dynamics of particles in random media which can explain a large variety of phenomena in statistical physics and condensed matter~\cite{ben}. Going back to finance, there is a wide consensus that volatility plays a key role in the dynamics of financial markets~\cite{bouchaudbook,lo,ding,stanley,black76,bollerslev,barndorff}. Among the most relevant statistical properties of the financial markets, volatility seems to be responsible for the observed clustering in price changes. That is, large fluctuations are commonly followed by other large fluctuations and similarly for small changes \cite{bouchaudbook}. Another feature is that, in clear contrast with price changes which show negligible autocorrelations, volatility autocorrelation is still significant for time lags longer than one year~\cite{bouchaudbook,bacry,bouchaud,lo,ding,lebaron}. Additionally, there exists the so-called leverage effect, {\it i.e.,} much shorter (few weeks) negative cross-correlation between current price change and future volatility~\cite{bouchaudbook,bouchaud-lev,black76,bollerslev}. Finally, several authors have argued that a good approximation for the volatility distribution can be given by the log-normal~\cite{bouchaudbook,montagna,stanley}. Another possible candidate to such distribution is provided by the so-called inverse Gaussian distribution~\cite{barndorff}.

All these observations have led us to consider the exponential Ornstein-Uhlenbeck (expOU) model, since we have seen that, among the simplest and classic SV models, the expOU model is able : (i) to describe simultaneously the observed long-range memory in volatility and the short one in leverage~\cite{comp,qf,qiu}, (ii) to provide a consistent stationary distribution for the volatility with data~\cite{qf,montagna,eisler}, (iii) it shows the same mean first-passage time profiles for the volatility as those of empirical daily data~\cite{extreme}, and finally (iv) it fairly reproduces the realized volatility having some degree of predictability in future return changes~\cite{eisler}. Our aim in this research is to take advantage of all this knowledge to study the European option price. We shall propose an approximate price for the option valid for realistic parameters that guarantees the empirical observations just mentioned. We therefore study the influence of these empirical observations in the price by designing an approximation procedure that provides an alternative price which is appropriate for a different range of parameters to those already published in Refs.~\cite{fouquebook} (see also the discussion at the end of Sect. III). Moreover, our present study goes much further than the simulation analysis already performed by one of us~\cite{pere2007}.

Before closing this introductory section, we should remark that the approximation procedure we use has been specially designed to tackle this kind of financial problems. However, the method is somewhat related to the Born-Oppenheimer approximation which, after its introduction in 1927 for the quantum mechanical treatment of molecules, has been widely used for a great variety of physical applications. In statistical mechanics the Born-Oppenheimer  approximation is specially suited for the so-called ``adiabatic elimination of fast variables'' (in our case the volatility)~\cite{risken}. A  simplified version of that approximation has been recently used to address some financial problems \cite{remer,biro}.

The paper is divided into five sections. Section~\ref{anal} presents the expOU model and shows its main statistical properties. In Section~\ref{optionproblem} we construct the equivalent martingale density and afterward derive the European call option price. Section~\ref{results} studies the price derived and compare it with the classic BS price. Conclusions are drawn in Section~\ref{conclusion} and some technical details are left to the Appendices.

\section{The Market Model: A Summary}
\label{anal}
As most SV models the correlated exponential Ornstein-Uhlenbeck (expOU) stochastic volatility model is a special kind of a two-dimensional diffusion process. We have thoroughly studied the statistical properties of the model in previous works with a quite satisfactory agreement with empirical data of real markets~\cite{pre-lev,comp,qf,eisler,extreme}. The model proposed by Scott~\cite{scott} in 1987 is furnished by a pair of It\^o stochastic differential equations:
\begin{eqnarray}
\frac{dS(t)}{S(t)}&=&\mu dt+ me^{Y(t)}dW_1(t)\label{price}\\
dY(t)&=&-\alpha Y(t)dt+kdW_2(t),\label{2d1}
\end{eqnarray}
where $S(t)$ is a financial price or the value of an index. The parameters $\alpha$, $m$, and $k$ are positive and nonrandom quantities and  $dW_i(t)=\xi_i(t)dt$ ($i=1,2$) are correlated Wiener processes, {\it i.e.}, $\xi_i(t)$ are zero-mean Gaussian white noise processes with cross correlations given by
\begin{equation}
\left\langle\xi_i(t)\xi_j(t')\right\rangle=\rho_{ij}\delta(t-t'),
\label{rho}
\end{equation}
where $\rho_{ii}=1$, $\rho_{ij}=\rho$ $(i\neq j, -1\leq\rho\leq 1)$. We remark that while the original Scott model~\cite{scott} had independent Brownian motions, it is important to allow for non-zero $\rho_{ij}$ to explain some empirical observations~\cite{fouquebook,fouque}.

From Eq.~(\ref{2d1}) we see that
\begin{equation}
Y(t)=Y_0e^{-\alpha (t-t_0)} + k\int^{t}_{t_0}e^{-\alpha(t-s)}dW_2(s),
\label{y}
\end{equation}
where we assume that the volatility process $Y(t)$ starts at certain initial time $t_0$ (which can be set equal to $0$) with a known value $Y(t_0)=Y_0$. The process $Y(t)$ is Gaussian with conditional first moment and variance given by
\begin{equation}
E[Y(t)|Y_0]=Y_0 e^{-\alpha(t-t_0)} \qquad \mbox{and} \qquad {\rm Var}[Y(t)|Y_0]=\frac{k^2}{2\alpha}\left(1-e^{-2\alpha(t-t_0)}\right).
\label{yave}
\end{equation}
In the stationary limit, $(t-t_0)\rightarrow \infty$, we have
\begin{equation}
E[Y(t)]=0\qquad \mbox{and} \qquad {\rm Var}[Y(t)]=\frac{k^2}{2\alpha}.
\label{yaves}
\end{equation}

In terms of the process $Y(t)$, the volatility is given by
\begin{equation}
\sigma(t)=me^{Y(t)}.
\label{sigma}
\end{equation}
Hence, the conditional probability density function (pdf) for the volatility is
\begin{equation}
p(\sigma,t|\sigma_0,0)=\frac{1}{\sigma\sqrt{2\pi\beta^2(1-e^{-2\alpha t})}}
\exp\left\{-\frac{[\ln(\sigma/m)-e^{-\alpha t}\ln(\sigma_0/m)]^2}{2\beta^2(1-
e^{-2\alpha t})}\right\},
\label{sigmapdf}
\end{equation}
and the stationary probability density reads
\begin{equation}
p_{st}(\sigma)=\frac{1}{\sigma\sqrt{2\pi\beta^2}}\exp\left\{-\ln^2(\sigma/m)/2
\beta^2\right\},
\label{statpdf}
\end{equation}
where
\begin{equation}
\beta^2=\frac{k^2}{2\alpha}.
\label{beta}
\end{equation}
This lognormal curve for the volatility is consistent with real data as it has been reported in several studies for different time lags~\cite{bouchaudbook,qf,stanley,eisler,montagna}.

Other interesting statistical properties of the model refer to the returns process 
given by
$$
dR(t)=dS(t)/S(t)-E[dS(t)/S(t)].
$$
We thus have the squared returns autocorrelation~\cite{qf}
\begin{equation}
\mbox{Corr}\left[dR(t)^2,dR(t+\tau)^2\right]=\frac{\exp[4\beta^2 e^{-\alpha\tau}]-1}{3e^{4\beta^2}-1},
\label{volcorfin}
\end{equation}
and the leverage effect (or return-volatility asymmetric correlation)~\cite{qf}
\begin{equation}
{\cal L}(\tau)=\frac{E\left[ dR(t)dR(t+\tau)^2\right]}{E\left[dR(t)^2\right]^2}=\frac{2\rho k}{m}\exp\left\{-\alpha\tau+2\beta^2(e^{-\alpha\tau}-3/4)\right\}H(\tau),
\label{leveragefin}
\end{equation}
where $H(\tau)$ is the Heaviside step function ($H(\tau)$=1 when $\tau\geq 0$, and $H(\tau)$=0 when $\tau<0$).

Both the volatility and the squared returns process~(\ref{volcorfin}) show a decay in their autocorrelation described by the desired cascade of exponentials which goes from very fast time scales to the slowest ones~\cite{lebaron}. From Eq.~(\ref{volcorfin}), one can obtain~\cite{qf}
$$
\mbox{Corr}\left[dR(t)^2,dR(t+\tau)^2\right]=\frac{1}{(3e^{4\beta^2}-1)}
\sum_{n=1}^{\infty}\frac{(4\beta^2)^n}{n!} e^{-n\alpha\tau},
$$
which goes as
\begin{equation}
\mbox{Corr}\left[dR(t)^2,dR(t+\tau)^2\right]\simeq  \frac{4\beta^2}{(3e^{4\beta^2}-1)} \,e^{-\alpha\tau} \qquad \mbox{for } \alpha\tau\gg 1.
\label{vauto}
\end{equation}
For $\alpha$ small enough, of the order of $10^{-3} {\rm days}^{-1}$, this may appear as a long range memory up to few trading years~\cite{qf}.

The model is similarly able to explain the rather swift leverage effect --up to few weeks-- which is well observed in actual financial markets. From Eq.~(\ref{leveragefin}), one can now obtain~\cite{qf}
\begin{equation}
{\cal L}(\tau)\simeq \frac{2\rho k}{m}\exp(\beta^2/2)e^{-k^2\tau} H(\tau) \qquad \mbox{for } \alpha\tau\ll 1,
\label{l2}
\end{equation}
and
\begin{equation}
{\cal L}(\tau)\simeq \frac{2\rho k}{m}\exp(-3\beta^2/2)e^{-\alpha\tau} H(\tau) \qquad \mbox{for } \alpha\tau\gg 1.
\label{l3}
\end{equation}
Note that the leverage amplitude in the latter regime is smaller than that over the shorter range (cf. Eqs.~(\ref{l2}) and~(\ref{l3})). Hence, the leverage asymptotics given by Eq.~(\ref{l3}) is negligible compared to the one provided by Eq.~(\ref{l2}). This result is consistent with real data showing that the leverage only persists up until few weeks and thus has a smaller range than the volatility autocorrelation. In this case, the model provides an exponential decay consistent with data if one takes characteristic time scale (given by the $k^2$ if we assume $k^2>\alpha$) between $10-100$ days and a moderate negative correlation between the two Wiener input noises $\rho\simeq-0.5$~\cite{qf}.

We remark that the range of parameters mentioned above is outside the values taken in Ref.~\cite{fouque}. In that case, Fouque {\it et al} assume that $\beta^2\sim 1$ (so that $k^2\sim \alpha$), thus implying that leverage and volatility autocorrelation have a very similar decay with a characteristic time of the same order (cf. Eqs.~(\ref{volcorfin})--(\ref{leveragefin}) and~(\ref{vauto})--(\ref{l2})).  Herein we consider a broader range for $\beta^2$~\cite{qf} in order to get a leverage decay faster than the volatility autocorrelation $k^2\gg\alpha$. Empirical observations have led us to consider the range where $\beta^2$ is much larger than one. In addition, Fouque {\it et al} also take $\alpha$ such as to provide a fast reversion to the normal level of volatility with a characteristic time scale $1/\alpha$ between one day and few weeks~\cite{fouque}. Even from this perspective their framework is far away from the one we here propose since we are assuming that reverting force is quite slow with $1/\alpha$ of the order of one or more years.

\section{The Option pricing problem}
\label{optionproblem}

In a previous paper~\cite{qf} we have addressed the question of the distribution of returns characterized by either the return probability density function (pdf) or by its characteristic function with a rather satisfactory fit for 20 day price return changes. We have there derived an approximate solution for the return pdf in terms of an expansion of Hermite polynomials with prefactors related to the skewness and the kurtosis of the expOU model. We want now to apply a similar analysis to the option pricing problem illustrated by the European call ${\cal C}(S,T)$. The payoff of this contract, that is, its final condition, is
\begin{equation}
{\cal C}(S,0)=\max[S(T)-K,0].
\label{final}
\end{equation}
where $S(T)$ is the underlying asset price at expiration date $T$ and $K$ is the strike price.

\subsection{The equivalent martingale measure}
\label{3a}

Following the standard approach to option pricing~\cite{hull,fouquebook},
we pass to an equivalent martingale (or risk-neutral) pricing measure $P^*$, under which the discounted price of traded securities are martingales. In particular, assuming a constant risk-free interest rate $r$:
$$
E^*\left[S(t)|S\right]=Se^{rt},
$$
where $S=S(0)$. Under $P^*$, the dynamics  (\ref{price})-(\ref{2d1}) of the expOU model are given by the following set of stochastic differential equations
\begin{eqnarray}
\frac{dS(t)}{S(t)}&=&r\,dt+me^{Y(t)}dW_1^*(t)\label{pricemart}\\
dY(t)&=&-\alpha Y(t)\,dt-k\Lambda(X,Y,t)\,dt+ k\,dW_2^*(t),
\label{2d}
\end{eqnarray}
where $(W_1^*,W_2^*)$ are $P^*$-Brownian motions with the same correlation structure as the original Brownian motions $(W_1,W_2)$.

In writing Eqs.~(\ref{pricemart})-(\ref{2d}) one has to include an arbitrary function $\Lambda(\cdot)$ called the log-volatility's market price of risk~\cite{hull,fouquebook}. This function quantifies the risk aversion sensitivity of the trader to the volatility uncertainty. Note that if there were no volatility fluctuations ({\it i.e.,} $k=0$) the function $\Lambda$ would vanish.

It does not seem possible to obtain the exact solution to this option problem and we will therefore search for approximate expressions but using different methods than those already proposed in the literature~\cite{fouque,fouquebook}. In order to specify the risk aversion function $\Lambda(X,Y,t)$ we first suppose that it depends solely on the volatility level and not on the return or the time. That is, $\Lambda(X,Y,t)=\Lambda(Y)$. Moreover, we further assume that this dependence is weak which amounts to $\Lambda(Y)$ being approximated by a linear function of $Y$:
\begin{equation}
\Lambda(Y)\approx\Lambda_0+\Lambda_1Y.
\label{llambda}
\end{equation}
We now define a new volatility-driving process
\begin{equation}
Z=Y+\frac{k\Lambda_0}{\bar{\alpha}},
\label{z}
\end{equation}
and the ``log return'' $X(t)$ by
\begin{equation}
X(t)=\ln[S(t)/S].
\label{martingale_return}
\end{equation}
Then Eqs.~(\ref{pricemart}) and~(\ref{2d}) turn into
\begin{eqnarray}
dX(t)&=&\left[r-\frac12 \bar{m}^2e^{2Z(t)}\right]dt+\bar{m}e^{Z(t)}dW_1^*(t)\label{pricemart1}\\
dZ(t)&=&-\bar{\alpha}Z(t)dt+ kdW_2^*(t),
\label{2d2}
\end{eqnarray}
where
\begin{equation}
\bar{m}=m\exp\left(-k\Lambda_0/\bar{\alpha}\right),\qquad \bar{\alpha}=\alpha+k\Lambda_1.
\label{mbar}
\end{equation}

We analyze the consequences of the introduction of a risk aversion function in the equivalent martingale process. Note first that the parameter $\bar{\alpha}$ which quantifies the memory range of the volatility (cf. Eq.~(\ref{vauto}) with $\alpha$ replaced by $\bar{\alpha}$) increases when $\Lambda_1>0$. Secondly, when $\Lambda_0>0$ the normal level of volatility shifts to lower values and the stationary volatility pdf has a smaller standard deviation (cf. Eqs.~(\ref{statpdf}),~(\ref{beta}) and~(\ref{mbar})).

\subsection{Approximate solution of the Fokker-Planck equation}

Let us denote by $p_2(x,z,t|0,z_0)$ the joint density of $(X_t,Z_t)$ under the equivalent martingale measure. This density obeys the following Fokker-Planck equation
\begin{eqnarray}
\frac{\partial p_2}{\partial t}=\bar{\alpha}\frac{\partial }{\partial z}(zp_2)+\frac{1}{2}k^2\frac{\partial^2p_2}{\partial z^2}
-\left(r-\frac12 e^{2z}\bar{m}^2\right)\frac{\partial p_2}{\partial x}+\frac12 e^{2z} \frac{\partial^2 p_2}{\partial x^2}+\rho k\bar{m}\frac{\partial^2}{\partial x\partial z}\left(e^zp_2\right),
\label{fpe2d}
\end{eqnarray}
with initial condition
\begin{equation}
p_2(x,z,0|0,z_0)=\delta(x)\delta(z-z_0).
\label{initial2}
\end{equation}

Even when the risk aversion is a linear function of the log-volatility, as is our case, we may only derive an approximate solution valid under particular values of the parameters of the model for a given market~\cite{qf}. Herein we will treat the case in which the ``vol of vol" $k$ is much greater than the normal level of volatility $\bar{m}$. In other words, we will assume that the parameter
\begin{equation}
\lambda\equiv\frac{k}{\bar{m}}\gg 1
\label{lambda}
\end{equation}
is large. Certainly this is at least the case of the Dow Jones Index daily data for which $\lambda\sim 10^2$~\cite{qf}.

We define a dimensionless time $t'$ and two scaling variables $u$ and $v$ by
\begin{equation}
t'=k^2t,\qquad u=\lambda x,\qquad v=\lambda z.
\label{scaling}
\end{equation}
The FPE (\ref{fpe2d}) then reads
\begin{eqnarray}
\frac{\partial p_2}{\partial t'}=\nu\frac{\partial }{\partial v}(vp_2)+
\frac{\lambda^2}{2}\frac{\partial^2p_2}{\partial v^2}&-&\frac{1}{\lambda}\Biggl(\frac{r}{\bar{m}^2}-\frac12 \bar{m}^2 e^{2v/\lambda}\Biggr)\frac{\partial p_2}{\partial u} \nonumber \\
&+&\frac{1}{2}e^{2v/\lambda}\frac{\partial^2p_2}{\partial u^2}+\lambda\rho\frac{\partial^2}{\partial u\partial v}\left(e^{v/\lambda}p_2\right),
\label{fpescaling0}
\end{eqnarray}
and
\begin{equation}
p_2(u,v,0|0,v_0)=\delta(u)\delta(v-v_0),
\label{initialscaling0}
\end{equation}
where
\begin{equation}
v_0=\lambda z_0, \qquad \mbox{and} \qquad \nu=1/(2\beta^2)=\bar{\alpha}/k^2.
\label{v0}
\end{equation}

We can write a more convenient equation in terms of the characteristic function defined as
\begin{equation}
\varphi(\omega_1,\omega_2,t'|0,v_0)=\int^{\infty}_{-\infty}e^{i\omega_1u}du
\int^{\infty}_{-\infty}e^{i\omega_2v}p_2(u,v,t'|0,v_0)dv.
\label{charf}
\end{equation}
This double Fourier transform turns the initial-value problem (\ref{fpescaling0})-(\ref{initialscaling0}) into
\begin{eqnarray}
\frac{\partial}{\partial t'}\varphi(\omega_1,\omega_2, t')=
&&-\nu\omega_2\frac{\partial}{\partial\omega_2}\varphi(\omega_1,\omega_2, t')-
\frac{1}{2}\lambda^2\omega_2^2\varphi(\omega_1,\omega_2, t') \nonumber\\
&&-\frac{i}{2\lambda}\omega_1\varphi(\omega_1,\omega_2-2i/\lambda, t')
+\frac{ir}{\lambda m^2}\omega_1\varphi(\omega_1,\omega_2, t')
\nonumber\\
&&-\frac{1}{2}\omega_1^2\varphi\left(\omega_1,\omega_2-2i/\lambda, t'\right)-\lambda\rho\omega_1\omega_2\varphi\left(\omega_1,\omega_2-i/\lambda, t'\right),
\label{cf2d}
\end{eqnarray}
with
\begin{equation}
\varphi(\omega_1,\omega_2,0|0,v_0)=e^{i\omega_2v_0}.
\label{cfinitial}
\end{equation}
Since the payoff for the European option depends on the price but not on the volatility, we only need to know the marginal characteristic function of the (martingale) return $X(t)$. This would imply solving Eq. (\ref{cf2d}) when $\omega_2=0$ and assume $\omega_1=\omega/\lambda$ (see Eq. (\ref{scaling})). Unfortunately, we cannot proceed in such a direct way and have to solve the two-dimensional problem when $\omega_2$ is small and $\lambda$ is large. This is done in Appendix~\ref{app1} where we prove the following approximate expression for the marginal characteristic function of the return:
\begin{equation}
\varphi(\omega/\lambda,\omega_2=0,t'|0,v_0)=\exp\left\{-C(\omega/\lambda,t')+\mbox{O}\left(1/\lambda^5\right)\right\}.
\label{charf1}
\end{equation}
where $C(\omega/\lambda,t')$ is given in Eq.~(\ref{mar5}) which, after recovering the original variables $t$ and $z_0$ (cf. Eq. (\ref{scaling})) and some reshuffling of terms, reads
\begin{eqnarray}
C(\omega/\lambda,\bar{\alpha} t)=i\mu(t)\omega+\frac{\bar{m}^2t}{2}\omega^2+\vartheta(t,z_0)\omega^2-i\rho \varsigma(t,z_0)\omega^3-\kappa(t)\omega^4+\mbox{O}(1/\lambda^5).
\label{mar6}
\end{eqnarray}
where
\begin{eqnarray}
&&\mu(t)=rt-\frac12 \bar{m}^2 t, \label{M}\\
&&\vartheta(t,z_0)=\frac{z_0}{\lambda^2\nu}\left(1-e^{-\bar{\alpha} t}\right),\label{V} \\
&&\varsigma(t,z_0)=\frac{1}{\lambda^3\nu^2}\left[\bar{\alpha} t-\left(1-e^{-\bar{\alpha} t}\right)\right]-\frac{z_0}{\lambda^3\nu^2}\left[\bar{\alpha} t e^{-\bar{\alpha} t}-\left(1-e^{-\bar{\alpha} t}\right)\right],\label{S} \\
&&\kappa(t)=\frac{1}{2\lambda^4\nu^3}\Biggl[\bar{\alpha} t+\frac12 \Bigl(1-e^{-2\bar{\alpha} t}\Bigr)-2\Bigl(1-e^{-\bar{\alpha} t}\Bigr)\Biggr]
+\frac{\rho^2}{2\lambda^4\nu^3}\left[\bar{\alpha} t-2\left(1-e^{-\bar{\alpha} t}\right)+\bar{\alpha} t e^{-\bar{\alpha} t}\right]. \nonumber\\
\label{K}
\end{eqnarray}

The approximate expression for the marginal characteristic function of the return is thus obtained by taking first terms of Taylor expansion of Eq.~(\ref{charf1}). We get
\begin{eqnarray}
&&\varphi(\omega/\lambda,\bar{\alpha}t)
=\exp\Biggl\{-\Bigl[i\omega\mu(t)+\bar{m}^2\omega^2t/2\Bigl]\Biggl\}\left[1
-\vartheta(t,z_0)\omega^2+i\rho\varsigma(t,z_0)\omega^3\right.\nonumber\\
&&\left.\qquad\qquad\qquad\qquad\qquad\qquad\qquad\qquad\qquad
+\left(\kappa(t)+\vartheta(t,z_0)^2/2\right)\omega^4+\mbox{O}(1/\lambda^5)\right].
\label{char}
\end{eqnarray}
The inverse Fourier transform of this expression yields the following approximate solution for the pdf of the (martingale) return:
\begin{eqnarray}
p(x,t|z_0)\simeq&&\frac{1}{\sqrt{2\pi \bar{m}^2t}}\exp\left[-\frac{(x-\mu)^2}{2\bar{m}^2t}\right]\left[1+\frac{\vartheta}{2\bar{m}^2t}H_2\left(\frac{x-\mu}{\sqrt{2\bar{m}^2t}}\right)\right.
\nonumber \\&&\left.+\frac{\rho\varsigma}{(2\bar{m}^2t)^{3/2}}H_3\left(\frac{x-\mu}{\sqrt{2\bar{m}^2t}}\right) +\frac{\kappa+\vartheta^2/2}{(2\bar{m}^2t)^2}H_4\left(\frac{x-\mu}{\sqrt{2\bar{m}^2t}}\right)\right],
\label{mar8}
\end{eqnarray}
where $\mu,\vartheta,\varsigma$, and $\kappa$ are given by Eqs.~(\ref{M})--(\ref{K}), and $H_n(\cdot)$ are the Hermite polynomials:
\begin{equation}
\int_{-\infty}^{\infty}x^n e^{-(ax)^2-iwx}dx=\frac{\sqrt{\pi}}{(-2ia)^na}e^{-w^2/4a^2} H_n(w/2a).
\label{hermite}
\end{equation}

As shown in Eqs.~(\ref{mbar}), we note that both the rate of mean-reversion of the log-volatility given by $\alpha$ and the normal level of volatility $m$ are rescaled depending on the own risk aversion of the agent. We also recall that the dependence on the original log-volatility $y_0=\ln(\sigma_0/m)$ is provided by (cf. Eq.~(\ref{z}))
\begin{equation}
z_0=y_0+\frac{k\Lambda_0}{\bar{\alpha}}.
\label{z0}
\end{equation}
Moreover, a further simplification can be achieved by averaging over the initial volatility with the zero-mean stationary (and Gaussian) distribution of the process $Z$ (cf. Eqs.~(\ref{2d2})). We can use same structure as given by Eq.~(\ref{mar8}) but replace some of the parameters involved by
\begin{eqnarray}
&&\vartheta(t,z_0)\rightarrow 0 \label{Vave} \\
&&\varsigma(t,z_0)\rightarrow \hat{\varsigma}(t)=\frac{1}{\lambda^3\nu^2}\left[\bar{\alpha} t-\left(1-e^{-\bar{\alpha} t}\right)\right].\label{Save}\\
&&\kappa(t)+\frac12\vartheta(t,z_0)^2\rightarrow\hat{\kappa}(t)=\frac{1}{2\lambda^4\nu^3}\Biggl\{\bar{\alpha} t-\Bigl(1-e^{-\bar{\alpha} t}\Bigr)+\rho^2\left[\bar{\alpha} t-2\left(1-e^{-\bar{\alpha} t}\right)+\bar{\alpha} t e^{-\bar{\alpha} t}\right]\Biggr\}. \nonumber\\
\label{Kave}
\end{eqnarray}

\begin{figure}
\epsfig{file=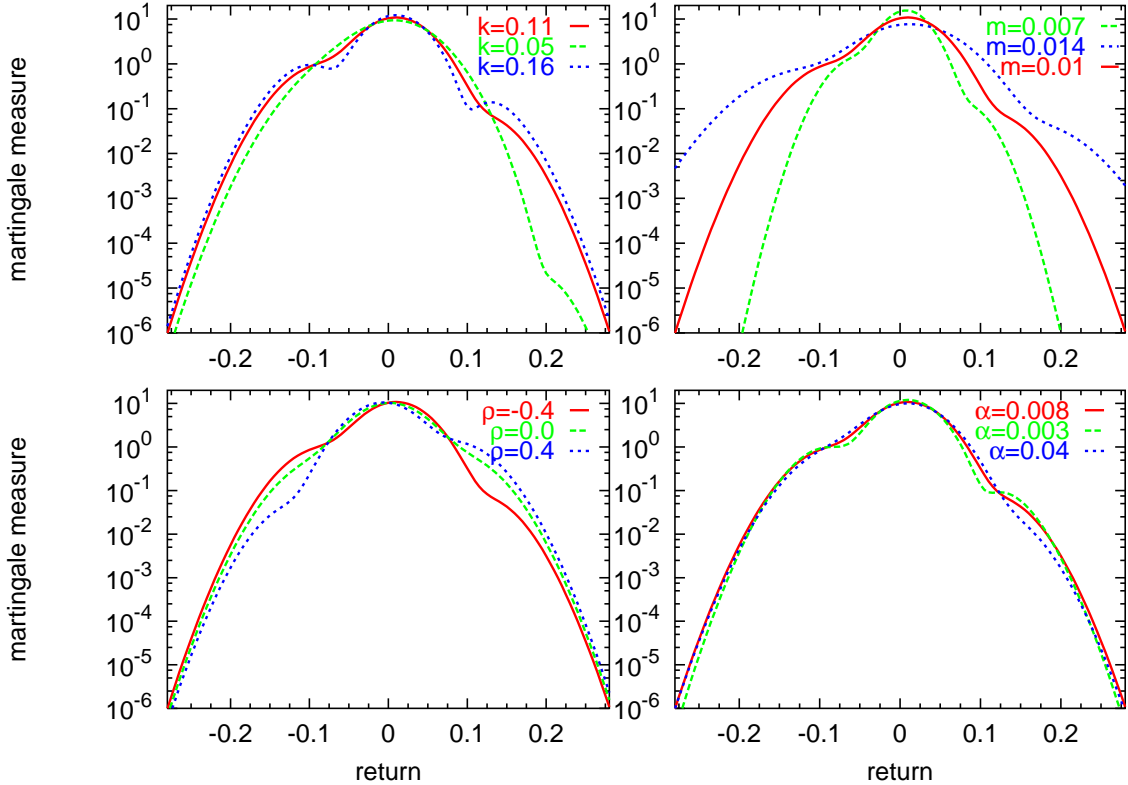,scale=1.2}
\caption{Risk-neutral return density~(\ref{mar8}) for $t=20$ days and with terms provided by Eqs.~(\ref{M})--(\ref{K}) when $z_0=0$ and assuming $\Lambda_0=10^{-3}$ and $\Lambda_1=10^{-3}$ (cf. Eq.~(\ref{llambda})). We depart from the parameters $m=10^{-2}\, {\rm day}^{-1/2}$, $\alpha=8\times 10^{-3}\, {\rm day}^{-1}$, $\rho=-0.4$ and $k=0.11\, {\rm day}^{-1/2}$ and slightly modify them in each of these plots .}
\label{fig-mart-parameters}
\end{figure}

In Fig.~\ref{fig-mart-parameters} we represent the approximate expression of the equivalent martingale measure as given by Eq.~(\ref{mar8}). The initial log-volatility is $z_0=0$ and the linear market price of risk is characterized by $\Lambda_0=\Lambda_1=0.001$ (cf. Eq.~(\ref{llambda})). We plot a set of distributions where we provide three different values of each parameter ($k,m,\rho$, and $\alpha$) defining the expOU model. We observe that the vol-of-vol $k$ mainly modifies the positive wing of the distribution. The normal level of volatility $m$ broadens the probability distribution with no difference between the two tails. The correlation between Wiener noises $\rho$ provide the observed negative skewness only if $\rho$ is negative so that this term should be taken into account if we want to include this effect to the corresponding option price. And finally, the long-range memory parameter $\alpha$ of the reverting force has little effect on the distribution profile. Hence overestimating (or underestimating) this quantity does not have great consequences in providing a good approximation of the risk-neutral distribution.

\subsection{The European Call price}

Our main goal is to study the European call option price. Once we have obtained our approximate solution for the equivalent martingale measure, the price can be computed in terms of the expected payoff~(\ref{final}) under our equivalent martingale measure~\cite{hull}. That is:
\begin{eqnarray}
{\cal C}(S,T,z_0)&=&e^{-rT}E\left[\left.\max(Se^{X(T)}-K,0)\right|X(0)=0,\bar{Z}(0)=z_0\right]\nonumber \\
&=&e^{-rT}\int_{-\infty}^{\infty} \max(Se^{x}-K,0)p(x,T|z_0) dx,
\label{call}
\end{eqnarray}
where the price at the expiration date $T$ is provided by the return path according to the relation $S(T)=S\exp(X(T))$ and $K$ is the strike price. As can be observed from the expansion~(\ref{mar8}), the computation of the approximate option price implies evaluating four integrals. The first one, $C_{BS}$, corresponding to the classic Black-Scholes price ({\it i.e.,} when the underlying process has a constant volatility given by $\bar{m}$):
\begin{equation}
{\cal C}_{BS}(S,T)=SN(d_1)-Ke^{-rT}N(d_2),
\label{bs}
\end{equation}
where
\begin{equation}
d_1=\frac{\ln S/K+(r+\bar{m}^2/2)T}{\sqrt{\bar{m}^2T}}, \qquad \mbox{and} \qquad d_2=\frac{\ln S/K+(r-\bar{m}^2/2)T}{\sqrt{\bar{m}^2T}}
\label{d}
\end{equation}
and $N(d)$ is the normal distribution
\begin{equation}
N(d)=\frac{1}{\sqrt{2\pi}}\int_{-\infty}^d e^{-x^2/2}dx.
\label{N}
\end{equation}

In the Appendix~\ref{app2} we evaluate the rest of terms. Summing them up, our approximate price for the European call option when underlying follows an expOU stochastic volatility process reads
\begin{eqnarray}
{\cal C}(S,T,z_0)={\cal C}_{BS}(S,T)+\left(\vartheta+\rho\varsigma+\kappa+\frac{\vartheta^2}{2}\right)SN(d_1)+\frac{Ke^{-rT}}{\sqrt{\bar{m}^2T}}N'(d_2)\left[\frac{\kappa+\vartheta^2/2}{2\bar{m}^2T}H_2\left(\frac{d_2}{\sqrt{2}}\right)\right.
\nonumber\\\left.
-\frac{\rho\varsigma+\kappa+\vartheta^2/2}{\sqrt{2\bar{m}^2T}}H_1\left(\frac{d_2}{\sqrt{2}}\right)+\vartheta+\rho\varsigma+\kappa+\vartheta^2/2\right],
\nonumber \\
\label{ct}
\end{eqnarray}
where ${\cal C}_{BS}$ is given by Eq.~(\ref{bs}), $d_1$ and $d_2$ are given by Eq.~(\ref{d}), $N'(x)=dN(x)/dx$ and $\vartheta$, $\varsigma$, and $\kappa$ are defined in Eqs.~(\ref{V})--(\ref{K}). It can be easily proven that the resulting price satisfies the so-called put-call parity
$$
{\cal C}(S,T,z_0)+Ke^{-rT}={\cal P}(S,T,z_0)+S
$$
where ${\cal P}$ is the price of the put option which is a derivative contract whose payoff is $\max(K-S,0)$~\cite{hull}. This relationship guarantees the absence of arbitrage, that is: the practice of taking advantage of a price differential between different assets without taking any risk. Finally when the initial volatility $z_0$ has been averaged out the price has same form as that of Eq.~(\ref{ct}) but using the parameters given by Eq.~(\ref{Vave}).

\section{Some results}
\label{results}

\begin{figure}
\epsfig{file=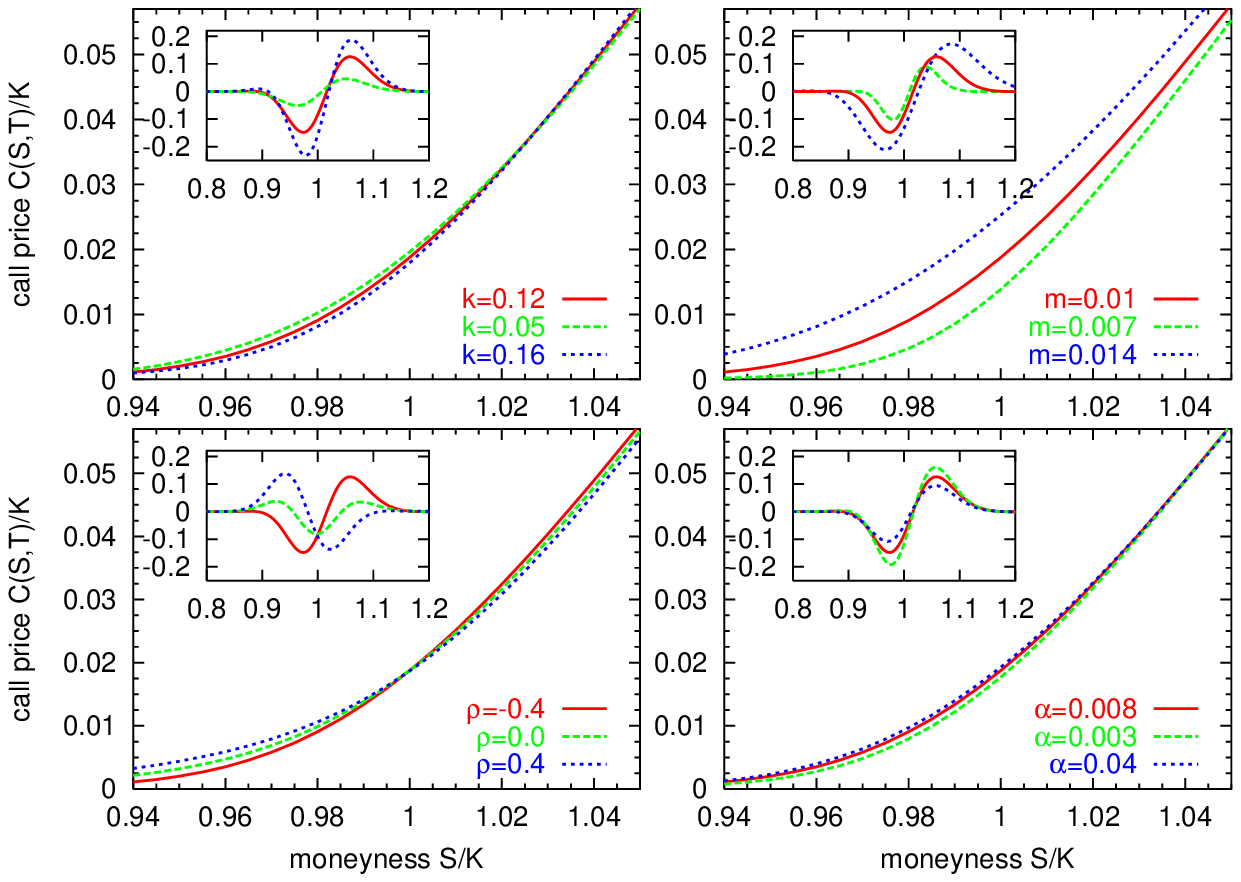,scale=1.2}
\caption{Normalized ${\cal C}/K$ call price~(\ref{ct}) as a function of the moneyness $S/K$ for $T=20$ days assuming $\Lambda_0=10^{-3}$ and $\Lambda_1=10^{-3}$ (cf. Eq.~(\ref{llambda})) with terms provided by Eqs.~(\ref{M})--(\ref{K}) when $z_0=0$. We depart from the parameters $m=10^{-2}\, {\rm day}^{-1/2}$, $\alpha=8\times 10^{-3}\, {\rm day}^{-1}$, $\rho=-0.4$ and $k=0.11\, {\rm day}^{-1/2}$ and slightly modify them in each of these plots.}
\label{fig-call-parameters}
\end{figure}

We will now analyze the call price given by Eq.~(\ref{ct}). In Fig.~\ref{fig-call-parameters} we show the effect of changing the parameters of the model on the resulting option price. As we did in Fig. \ref{fig-mart-parameters} each plot slightly modifies only one of the four parameters while the other three are kept constant although taking realistic values. Looking at Fig.~\ref{fig-call-parameters} we may say that the call price is highly sensitive to the normal level $m$. This is, however, not surprising since an extreme sensitivity to volatility --specially around moneyness $S/K=1$-- is a characteristic feature of the classic BS price~\cite{hull}. As an overall statement we may say that having large values of any parameter implies (except for $\rho$) a more expensive option.

We also compare our price, Eq.~(\ref{ct}), with the BS price, ${\cal C}_{BS}$, given by Eq.~(\ref{bs}) in which the  volatility is constant. The insets in Fig.~\ref{fig-call-parameters} represent the difference ${\cal C}-{\cal C}_{BS}$ and we there observe a distinct behavior depending on whether we have moneyness smaller than one (in-the-money option) or larger than one (out-the-money option). In general, our in-the-money calls are cheaper than the BS ones while the out-the-money calls become more expensive than the BS ones. This is however true as long as $\rho$ is negative since otherwise we would have the opposite effect. Let us recall that $\rho$ should be negative because of the negative skewness in the return distribution and also due to the negative return-volatility asymmetric correlation (leverage effect), both properties observed in actual markets. The profile of the call difference ${\cal C}-{\cal C}_{BS}$ that we have obtained is indeed consistent with the observed one such as in Ref.~\cite{Pochart} where, although based on a different option pricing method and on a different market model as well, the correlation coefficient, $\rho$, is studied with special attention. We also note that the impact on the call price of changing the reverting force $\alpha$ is much smaller than that of changing the vol-of-vol $k$ or the normal level $m$.

\begin{figure}
\epsfig{file=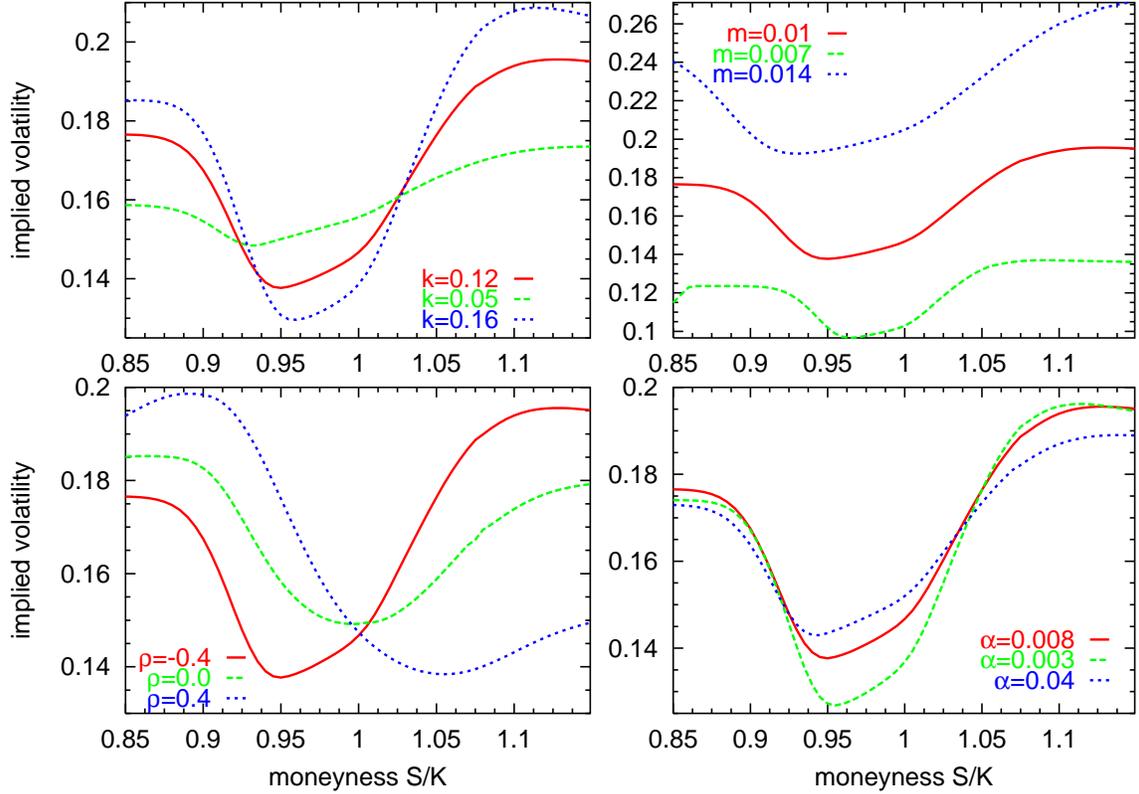,scale=1.2}
\caption{Implied volatility (in yearly units) as a function of the moneyness $S/K$ for $T=20$ days assuming $\Lambda_0=10^{-3}$ and $\Lambda_1=10^{-3}$ (cf. Eq.~(\ref{llambda})) with terms provided by Eqs.~(\ref{M})--(\ref{K}) when $z_0=0$. We depart from the parameters $m=10^{-2}\, {\rm day}^{-1/2}$, $\alpha=8\times 10^{-3}\, {\rm day}^{-1}$, $\rho=-0.4$ and $k=0.11\, {\rm day}^{-1/2}$ and slightly modify them in each of these plots.}
\label{fig-smile-parameters}
\end{figure}

We now go one step further and study the implied volatility $\sigma_{\rm i}$. This is the volatility that the classic BS formula should adopt if we require that ${\cal C}_{BS}(\sigma_{\rm i})={\cal C}$. We evaluate $\sigma_{\rm i}$  numerically in terms of the moneyness and for an identical set of parameters than those presented in previous figures. We observe that the vol-of-vol, $k$, and the long range memory parameter, $\alpha$, have both a rather similar effect although the steepest profile corresponds in one case to smaller values (the case of $\alpha$) while in the other case it corresponds to larger values (the case of $k$). In contrast, the normal level of volatility $m$ simply shifts the implied volatility profile to lower or higher volatility levels it keeps the same form of the profile. The smile effect is smirked in one side or in the other depending on the sign of $\rho$. The rest of the plots studying parameters $k,m$, and $\alpha$ are conditioned to this sign and for all of them we take $\rho=-0.4$.

\begin{figure}
\epsfig{file=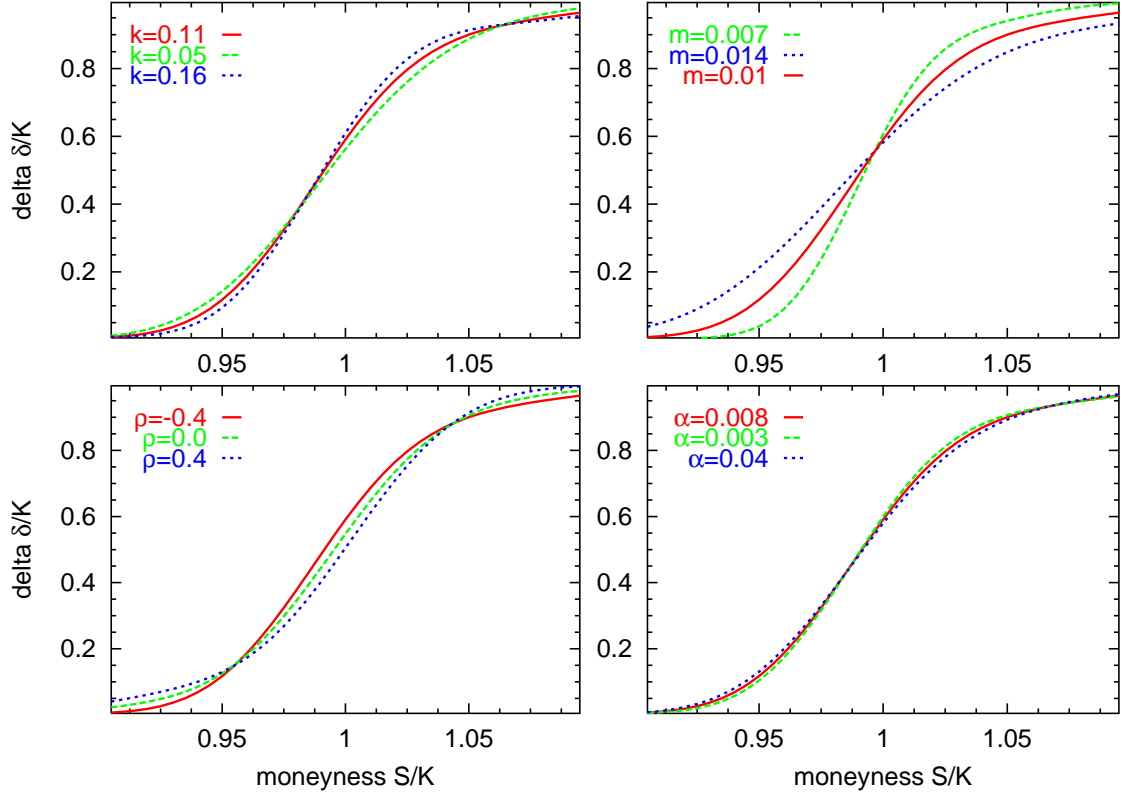,scale=1.2}
\caption{Delta hedging~(\ref{delta}) divided by strike $K$ as a function of the moneyness $S/K$ for $T=20$ days assuming $\Lambda_0=10^{-3}$ and $\Lambda_1=10^{-3}$ (cf. Eq.~(\ref{llambda})) with terms provided by Eqs.~(\ref{M})--(\ref{K}) when $z_0=0$. We depart from the parameters $m=10^{-2}\, {\rm day}^{-1/2}$, $\alpha=8\times 10^{-3}\, {\rm day}^{-1}$, $\rho=-0.4$ and $k=0.11\, {\rm day}^{-1/2}$ and slightly modify them in each of these plots.}
\label{fig-delta-parameters}
\end{figure}

\begin{figure}
\epsfig{file=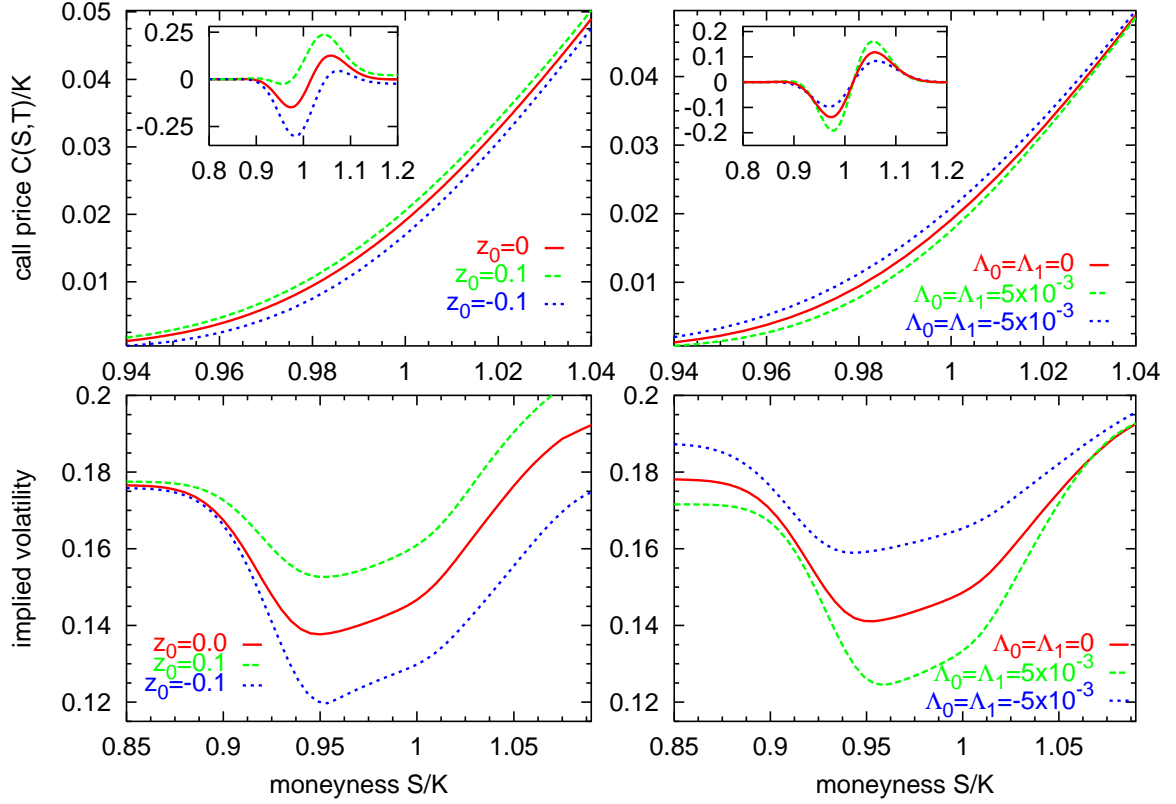,scale=1.2}
\caption{Call price~(\ref{call}) and implied volatility (in yearly units) as a function of the moneyness $S/K$ for $T=20$ days. Left column studies the effects of a non-zero initial volatility assuming $\Lambda_0=10^{-3}$ and $\Lambda_1=10^{-3}$. Right column shows those caused by changing the constant involved in the risk aversion function~(\ref{llambda}) when $z_0=0$. The rest of parameters are $m=10^{-2}\, {\rm day}^{-1/2}$, $\alpha=8\times 10^{-3}\, {\rm day}^{-1}$, $\rho=-0.4$ and $k=0.11\, {\rm day}^{-1/2}$.}
\label{fig-vol-risk-parameters}
\end{figure}

We can also provide an analytical expression for the delta hedging~\cite{hull,bouchaudbook}. This is a crucial magnitude since it specifies the number of shares per call to hold in order to remove risk of underlying asset price fluctuations (but not volatility fluctuations) from the portfolio. From Eq.~(\ref{ct}), it is straightforward to obtain
\begin{eqnarray}
\delta=\frac{\partial {\cal C}}{\partial S}=
\left(1+\vartheta+\rho\varsigma+\kappa+\vartheta^2/2\right)N(d_1)+
\frac{Ke^{-rT}}{S\sqrt{\bar{m}^2T}}N'(d_2)\left[-\frac{\kappa+\vartheta^2/2}{(2\bar{m}^2T)^{3/2}}H_3\left(\frac{d_2}{\sqrt{2}}\right)\right.
\nonumber\\
\left.
+\frac{\rho\varsigma+\kappa+\vartheta^2/2}{2\bar{m}^2T}H_2\left(\frac{d_2}{\sqrt{2}}\right)-\frac{\vartheta+\rho\varsigma+\kappa+\vartheta^2/2}{\sqrt{2\bar{m}^2T}}H_1\left(\frac{d_2}{\sqrt{2}}\right)+\vartheta+\rho\varsigma+\kappa+\vartheta^2/2\right].\nonumber\\
\label{delta}
\end{eqnarray}
Figure~\ref{fig-delta-parameters} thus provides the same set of plots as those of the previous cases. The long range memory parameter $\alpha$ has little effect in the delta hedging although $\delta$ has a higher sensitivity to the rest of parameters. Smaller values of the normal level of volatility $m$ make steeper the delta hedging profile. The correlation $\rho$ and the vol-of-vol $k$ have a non-trivial effect depending on the moneyness.

In Fig.~\ref{fig-vol-risk-parameters} we analyze the effects on the call price of the initial volatility and the risk aversion. Until now we have taken $z_0=0$ but now we can consider other possible values. As expected, we observe in Fig.~\ref{fig-vol-risk-parameters} that the call becomes more expensive if one takes an initial volatility greater than the normal level and cheaper in the opposite case (see also discussion in Section~\ref{3a}). We also look at the risk aversion terms provided by Eq.~(\ref{llambda}). Negative terms would thus correspond to a more expensive call while positive terms imply having a cheaper option since the agent is less risk averse.

\begin{figure}
\epsfig{file=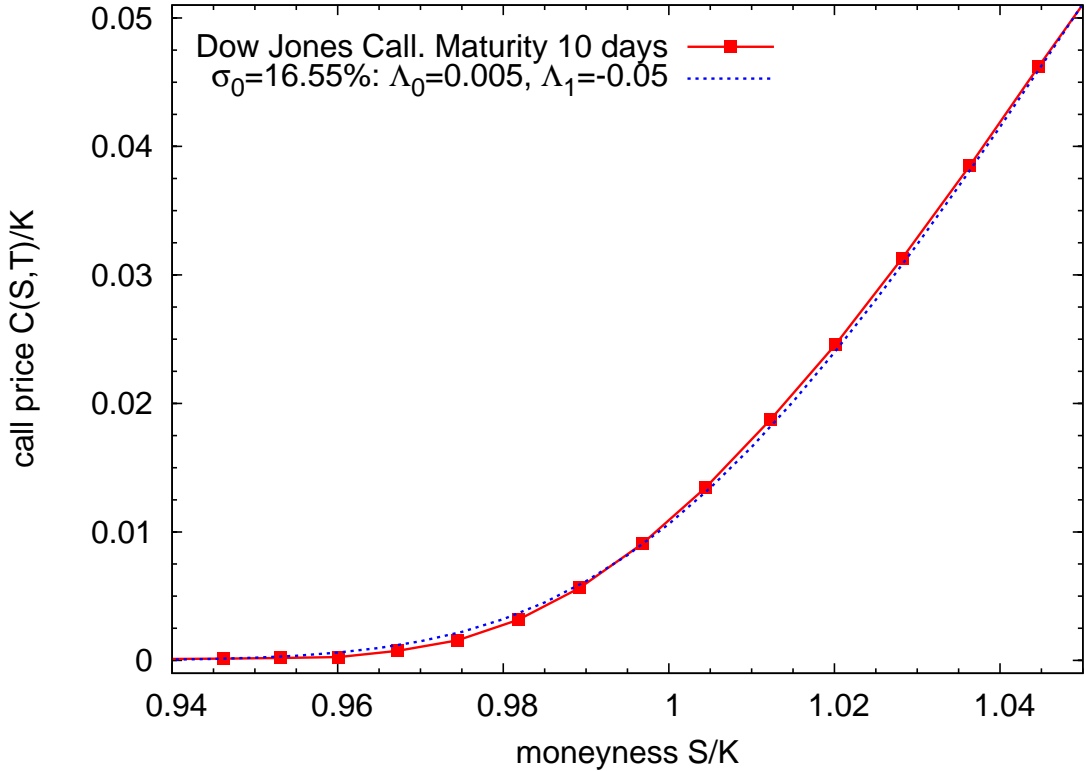,scale=1.2}
\caption{Call price as a function of the moneyness $S/K$ for $T=10$ days. Points represent the empirical call option prices on the Dow Jones Index (DJX) at a precise date (May 2, 2008) and with maturity on May 16, 2008. Dashed line takes a call price~(\ref{ct}) fit having fixed the model parameters estimated from historical data and with the initial volatility assumed to be the CBOE DJIA Volatility Index (VDX) and the current interest rate ratio $r=2\%$. The curve thus provides a fit with proper risk aversion parameters $\Lambda_0$ and $\Lambda_1$.}
\label{fig-impliedrsk}
\end{figure}

Before concluding this section, we question ourselves whether the risk averse parameters $\Lambda_0$ and $\Lambda_1$ can be in some way or another inferred from empirical data. These parameters give reason of the risk averse perception of the investors and depend on the situation of the market at a particular time. Just as an illustrative example, we can look at the European option contracts (DJX) on the Dow Jones Industrial Average index traded in the Chicago Board Exchange (CBOE) at a given date (2nd of May 2008) and for a particular maturity (16th of May 2008, 10 trading days ahead). We next assume that the call price is given by Eq.~(\ref{ct}) with model parameters to be those estimated historical data ($m=10^{-2}\, {\rm day}^{-1/2}$, $\alpha=8\times 10^{-3}\, {\rm day}^{-1}$, $\rho=-0.4$ and $k=0.11\, {\rm day}^{-1/2}$), take the US current risk-free interest ratio (annual rate $r=2\%$), and finally consider initial volatility to be the CBOE Volatility Index (VDX) on the 2nd of May 2008 (annual rate $\sigma_0=16.55\%$ from which we get $y_0=\ln(\sigma_0/m)$). The estimation on $\sigma_0$ can be more or less sophisticated~\cite{eisler} but it is rather usual to assume the VDX since this is designed to reflect investors' consensus on current volatility level. Figure~\ref{fig-impliedrsk} therefore presents a rather satisfactory fit on the empirical option contracts over different moneyness $S/K$ by solely modifying the risk parameters $\Lambda_0$ and $\Lambda_1$. The error in the fitting is rather small being of the same order or even smaller than the tick size of the traded option. In this way and thanks to the fact that the parameters of the model are obtained from historical Dow Jones data, we can obtain the implied risk aversion parameters which typically fluctuates in time as investors' perception changes.

\section{Concluding remarks}
\label{conclusion}

The main goal of this paper has been to study the effects on option pricing of several well-known properties of financial markets. These properties include the long-range memory of the volatility, the short-range memory of the leverage effect, the negative skewness and the kurtosis. The analysis is based on a market model that satisfies these properties which, in turn, can be easily identified through the parameters of the model. In this way we provide a different and more complete analysis on option pricing than that we had presented some time ago in which the effects of non-ideal market conditions such as fat tails and a small relaxation were taken into account~\cite{nonideal,correlated}.

We have derived an approximated European  call option prices when the volatility of the underlying price is random and it is described by the exponential Ornstein-Uhlenbeck process. The solution has been obtained by an approximation procedure based on a partial expansion of the characteristic function under the risk-neutral pricing measure.

The call price obtained is valid for a range of parameters different than those of a previous study on the subject~\cite{fouque}. In that work Fouque {\it et el} assumed that the reversion toward the normal level of volatility is fast. In other words, the parameter $\alpha$ is large and the characteristic time scale for reversion, $1/\alpha$, is of the order of few days. Fouque {\it et el} also considered that $\beta^2=k^2/2\alpha\sim 1$ which implies that the return-volatility asymmetric correlation ({\it i.e.,} the leverage effect) should have a characteristic time comparable to that of the volatility autocorrelation. Mostly based on the Dow Jones daily index data~\cite{qf,eisler}, we have considered a rather different situation where the fast parameter is not the reverting force $\alpha$ but the vol-of-vol $k$. In this way, we have singled out these two market memories thus allowing for a leverage effect during a time-lapse of few weeks and a persistent volatility autocorrelation larger than one year. These properties are consistent with empirical observations on the Dow-Jones index~\cite{qf}.

Under these circumstances we have constructed an approximate option price where risk aversion is assumed to be a linear function of the logarithm of the volatility. This approximation contains corrections in the variance, the skewness and the kurtosis, all of these corrections in terms of Hermite polynomials. This constitutes a tangible step forward with respect to other approaches which use the Heston stochastic volatility model~\cite{biro,remer} but only consider zeroth-order corrections. Our approach to the martingale measure albeit being more complete than those taken in Refs.~\cite{biro,remer} it is still able to provide an analytical expression for the call price and the subsequent Greeks.

Summarizing, we have studied the call price and its implied volatility and observed that the correlation $\rho$ between the two Wiener input noises plays a crucial role. The behavior of the call can greatly change depending on the sign of $\rho$ which confirms the findings of the previous work of Pochart and Bouchaud~\cite{Pochart} and many others. We have therefore focused on a negative value of $\rho$ that is consistent with empirical observations of the leverage effect. Keeping $\rho$ constant, moderate values of risk aversion, and a maturity time of the order of few weeks, we have also observed that the vol-of-vol, $k$, and the normal level of volatility, $m$, have both an important impact to the option price. This appears in clear contrast with the very little effect of not having a reliable estimation of the rate of mean-reversion of the volatility quantified by the parameter $\alpha$.

\appendix

\section{Approximate solution of the characteristic function}
\label{app1}

We start from Eqs. (\ref{cf2d})-(\ref{cfinitial}) and look for an approximate expression of the joint distribution $\varphi(\omega_1,\omega_2, t')$ valid for large values of $\lambda$. We also note that the marginal characteristic function of the (martingale) return can be obtained from the joint characteristic function by setting $\omega_2=0$. Therefore, we will look for a solution to the problem (\ref{cf2d})-(\ref{cfinitial}) that for small values of $\omega_2$ takes the form:
\begin{equation}
\varphi(\omega_1,\omega_2, t')=\exp\left\{-\left[A(\omega_1, t')\omega_2^2
+B(\omega_1, t')\omega_2+C(\omega_1, t')+{\rm O}(\omega_2^3,1/\lambda^2)\right]
\right\}.
\label{solutioncf2}
\end{equation}
Substituting this into Eq. (\ref{cf2d}) yields
\begin{eqnarray}
\dot{A}\omega_2^2+\dot{B}\omega_2+\dot{C}&=&-2\nu\omega_2^2A-\nu\omega_2B+\frac{\lambda^2}{2}\omega_2^2-\frac{i\omega_1}{2\lambda}+\frac{ir\omega_1}{m^2\lambda}+\frac12 \omega_1^2\left(1+4\frac{i\omega_2}{\lambda}A+\frac{2i}{\lambda}B\right)\nonumber\\
&&+\lambda\rho\omega_1\omega_2\left(1+\frac{2i\omega_2}{\lambda}A+\frac{i}{\lambda}B\right)+{\rm O}(1/\lambda^2),
\label{solutioncf3}
\end{eqnarray}
where the dot denotes a time derivative.

Collecting the quadratic terms in $\omega_2$, we get
$$
\dot{A}=-2\left(\nu-i\rho\omega_1\right)A+\frac{\lambda^2}{2}.
$$
The solution to this equation with the initial condition $A( t'=0)=0$ is
\begin{equation}
A(\omega_1, t')=\frac{\lambda^2}{4\gamma(\omega_1)}\left[1-e^{-2\gamma(\omega_1) t'}\right]
\label{AA}
\end{equation}
where
\begin{equation}
\gamma(\omega_1)=\nu-i\rho\omega_1.
\label{ghat}
\end{equation}
We now take the linear terms in $\omega_2$:
$$
\dot{B}=-\gamma B+\frac{2i\omega_1^2}{\lambda}A+\lambda\rho\omega_1.
$$
The solution to this equation with the initial condition $B( t'=0)=-iv_0$ is
\begin{equation}
B(\omega_1, t')=-iv_0e^{-\gamma t'}+\frac{\lambda\rho\omega_1}{\gamma}\left(1-e^{-\gamma t'}\right)+\frac{i\lambda\omega_1^2}{2\gamma^2}\left(1-e^{-\gamma t'}\right)^2
\label{BB}
\end{equation}
where $\gamma=\gamma(\omega_1)$ given by Eq.~(\ref{ghat}). Finally, the terms independent of $\omega_2$ yield
$$
\dot{C}=\frac12\omega_1^2+\frac{i\omega_1}{\lambda}\left(\frac{r}{\bar{m}^2}-\frac12\right)+\frac{i\omega_1^2}{\lambda}B
$$
with $C( t'=0)=0$. Therefore,
\begin{eqnarray}
C(\omega_1, t')&=&\left(\frac{r}{\bar{m}^2}-\frac12\right)\frac{i\omega_1}{\lambda} t'+\frac12\omega_1^2 t'+\frac{v_0\omega_1^2}{\lambda\gamma}\left(1-e^{-\gamma t'}\right)+\frac{i\rho\omega_1^3}{\gamma}\left[ t'-\frac{1}{\gamma}\left(1-e^{-\gamma t'}\right)\right]
\nonumber\\
&&\qquad\qquad
-\frac{\omega_1^4}{2\gamma^2}\left[ t'+\frac{1}{2\gamma}\left(1-e^{-2\gamma t'}\right)-\frac{2}{\gamma}\left(1-e^{-\gamma t'}\right)\right].
\label{CC}
\end{eqnarray}
We have thus obtained all the terms in the two-dimensional characteristic function determined by Eq.~(\ref{cf2d}).

However, that we only need to know the marginal characteristic function of the return $X(t)$ which implies that we only have to solve the equation when $\omega_2=0$ and assume $\omega_1=\omega/\lambda$. From Eq.~(\ref{solutioncf2}) we have the approximation
\begin{equation}
\varphi(\omega/\lambda,0, t')=\exp\left\{-C(\omega/\lambda, t')\right\},
\label{cf}
\end{equation}
where
\begin{eqnarray}
C(\omega/\lambda, t')&=&\left(\frac{r}{\bar{m}^2}-\frac12\right)\frac{i\omega}{\lambda^2} t'+\frac{\omega^2}{2\lambda^2} t'+\frac{v_0\omega^2}{\lambda^3\gamma}\left(1-e^{-\gamma t'}\right)+\frac{i\omega^2\eta}{\lambda^3\gamma}\left[ t'-\frac{1}{\gamma}\left(1-e^{-\gamma t'}\right)\right]
\nonumber\\
&&\qquad\qquad -\frac{\omega^4}{2\lambda^4\gamma^2}\left[ t'-\frac{1}{2\gamma}\left(1-e^{-2\gamma t'}\right)-\frac{2}{\gamma}\left(1-e^{-\gamma t'}\right)\right].
\label{mar1}
\end{eqnarray}
Note that there is an extra parameter involved :
$$
\gamma(\omega_1=\omega/\lambda)=\nu-\frac{i\rho\omega}{\lambda}
$$
that depends on $\lambda$ and it leads us to write $C(\omega/\lambda, t')$ in a somewhat more compact form (cf. Eqs.~(\ref{ghat})). Indeed, the first term we have to reconsider is
\begin{eqnarray}
\frac{1}{\lambda^3\gamma}\left(1-e^{-\gamma t'}\right)
=\frac{1}{\lambda^3\nu}\left[(1-e^{-\nu t'})-\frac{i\rho\omega}{\lambda\nu}\left(\nu t' e^{-\nu t'}-(1-e^{-\nu t'})\right)\right]+\mbox{O}(1/\lambda^5)
\nonumber\\
=\frac{1}{\lambda^3\nu}a( t')-\frac{i\rho\omega}{\lambda^4\nu^2}\left(b( t')-a( t')\right)+\mbox{O}(1/\lambda^5);
\label{mar2}
\end{eqnarray}
the second one reads
\begin{eqnarray}
&&\frac{1}{\lambda^3\gamma}\left[ t'-\frac{1}{\gamma}\left(1-e^{-\gamma t'}\right)\right]
\nonumber\\&&\qquad=\frac{1}{\lambda^3\nu^2}\left[\nu t'-\left(1-e^{-\nu t'}\right)+\frac{i\rho\omega}{\lambda\nu}\left(\nu t'(1+e^{-\nu t'})-2(1-e^{-\nu t'})\right)\right]+\mbox{O}(1/\lambda^5)
\nonumber\\&&\qquad=\frac{1}{\lambda^3\nu^2}\left[\left(1+\frac{i\rho\omega}{\lambda\nu}\right)\nu t'-\left(1+\frac{2i\rho\omega}{\lambda\nu}\right)a( t')+\frac{i\rho\omega}{\lambda\nu}b( t')\right]+\mbox{O}(1/\lambda^5);
\label{mar3}
\end{eqnarray}
while the third one is
\begin{eqnarray}
\frac{1}{\lambda^4\gamma^2}\left[ t'-\frac{1}{2\gamma}\left(1-e^{-2\gamma t'}\right)-\frac{2}{\gamma}\left(1-e^{-\gamma t'}\right)\right]&=&\frac{1}{\lambda^4\nu^3}\left[\nu t'-\frac{1}{2}\left(1-e^{-2\nu t'}\right)-2\left(1-e^{-\nu t'}\right)\right]\nonumber\\
&=& \frac{1}{\lambda^4\nu^3}\left[\nu t'-\frac12 a(2 t')-2a( t')\right]+\mbox{O}(1/\lambda^5),\nonumber\\
\label{mar4}
\end{eqnarray}
where
$$
a( t')= 1-e^{-\nu t'}, \qquad b( t')= \nu t' e^{-\nu t'} \label{bb}.
$$
All these expressions serve us to study the terms included in $C(\omega, t')$ given by Eq.~(\ref{mar1}) up to order $1/\lambda^4$. We sum up the contributions~(\ref{mar2})--(\ref{mar4}) and obtain
\begin{eqnarray}
C(\omega/\lambda, t')=\left(\frac{r}{\bar{m}^2}-\frac12\right)\frac{i\omega}{\lambda^2} t'+\frac{\omega^2}{2\lambda^2} t'+\frac{v_0\omega^2}{\lambda^3\nu}a( t')-\frac{i\rho v_0\omega^3}{\lambda^4\nu^2}\left[b( t')-a( t')\right]\nonumber\\
+\frac{i\rho\omega^3}{\lambda^3\nu^2}\left[\left(1+\frac{i\rho\omega}{\lambda\nu}\right)\nu t'-\left(1+\frac{2i\rho\omega}{\lambda\nu}\right)a( t')+\frac{i\rho\omega}{\lambda\nu}b( t')\right]
\nonumber\\
-\frac{\omega^4}{2\lambda^4\nu^3}\left[\nu t'+\frac12 a(2 t')-2a( t')\right]+\mbox{O}(1/\lambda^5).
\label{mar5}
\end{eqnarray}
We finally rearrange this expression taking into account the order of $\omega$. The final result is shown in Eq.~(\ref{mar6}) of the main text.

\section{Derivation of the European call option}
\label{app2}

We perform the average given by Eq.~(\ref{call}). Due to the fact that we have four contributions in Eq.~(\ref{mar8}), we will also obtain four terms for the option price. We decompose them as follows
\begin{equation}
{\cal C}(S,T,z_0)={\cal C}_{BS}(S,T)+\vartheta(T,z_0){\cal C}_0(S,T)+
\rho\varsigma(T,z_0) {\cal C}_1(S,T)+\left[\kappa(T)+\frac12\vartheta(T,z_0)^2\right]{\cal C}_2(S,T),
\label{c}
\end{equation}
where first term ${\cal C}_{BS}$ corresponds to the Black-Scholes price ({\it i.e.,} when underlying process has a constant volatility given by $\bar{m}$). The following terms --containing several corrections in the volatility, the skewness and kurtosis-- can be easily derived if one considers
\begin{equation}
e^{-a^2}H_n(-a)=\frac{d^n}{da^n}e^{-a^2},
\label{hn}
\end{equation}
where $H_n(\cdot)$ are the Hermite polynomials. The first term due to a non constant volatility is
\begin{eqnarray*}
{\cal C}_0(S,T)=\frac{e^{-rT}}{2\bar{m}^2t}\int_{-\infty}^\infty H_2\left(\frac{x-\mu}{\sqrt{2\bar{m}^2T}}\right) \frac{1}{\sqrt{2\pi \bar{m}^2T}}\exp\left[-\frac{(x-\mu)^2}{2\bar{m}^2T}\right]\max(S e^x-K,0)dx
\end{eqnarray*}
and taking into account Eq.~(\ref{hn}) it becomes
\begin{eqnarray*}
{\cal C}_0(S,T)=\frac{e^{-rT}}{2\bar{m}^2t\sqrt{\pi}}\int_{\frac{\ln(K/S)-\mu}{\sqrt{2\bar{m}^2T}}}^\infty \left(S e^{\mu+\sqrt{\bar{m}^2T}a}-K\right) \frac{d^2}{da^2}e^{-a^2} da,
\end{eqnarray*}
which, after some manipulations, finally reads
\begin{eqnarray}
{\cal C}_0(S,T)=SN(d_1)+\frac{Ke^{-rT}}{\sqrt{\bar{m}^2T}}N'(d_2),
\label{c12}
\end{eqnarray}
where $N'(x)=dN(x)/dx$ and $d_1$ and $d_2$ are defined in Eq.~(\ref{d}).

The second term of our calculation is
\begin{eqnarray*}
{\cal C}_1(S,T)=\frac{e^{-rT}}{(2\bar{m}^2t)^{3/2}}\int_{-\infty}^\infty H_3\left(\frac{x-\mu}{\sqrt{2\bar{m}^2T}}\right) \frac{1}{\sqrt{2\pi \bar{m}^2T}}\exp\left[-\frac{(x-\mu)^2}{2\bar{m}^2T}\right]\max(S e^x-K,0)dx.
\end{eqnarray*}
Again taking into account Eq.~(\ref{hn}), we have
\begin{eqnarray*}
{\cal C}_1(S,T)=\frac{e^{-rT}}{(2\bar{m}^2T)^{3/2}}\int_{\frac{\ln(K/S)-\mu}{\sqrt{2\bar{m}^2T}}}^\infty (S e^{\mu+\sqrt{\bar{m}^2T}a}-K) \frac{d^3}{da^3}e^{-a^2} da,
\end{eqnarray*}
which, after some simple algebra, yields
\begin{eqnarray}
{\cal C}_1(S,T)=
SN(d_1)-\frac{Ke^{-rT}}{\sqrt{\bar{m}^2T}}N'(d_2)\left[\frac{1}{\sqrt{2\bar{m}^2T}}H_1\left(d_2/\sqrt{2}\right)-1\right].
\label{c22}
\end{eqnarray}
Note that correlation $\rho$ between the two Brownian noise sources determines the sign and the strength of this term. Obviously if there is no correlation this term disappears (cf. Eq.~(\ref{c})).

The third and last piece of our option price reads
\begin{eqnarray*}
{\cal C}_2(S,T)=\frac{e^{-rT}}{(2\bar{m}^2t)^{2}}\int_{-\infty}^\infty H_4\left(\frac{x-\mu}{\sqrt{2\bar{m}^2T}}\right)\frac{1}{\sqrt{2\pi \bar{m}^2T}}\exp\left[-\frac{(x-\mu)^2}{2\bar{m}^2T}\right]\max(S e^x-K,0)dx
\end{eqnarray*}
and, after using Eq.~(\ref{hn}), we get
\begin{eqnarray*}
{\cal C}_2(S,T)=\frac{e^{-rT}}{(2\bar{m}^2T)^{2}}\int_{\frac{\ln(K/S)-\mu}{\sqrt{2\bar{m}^2T}}}^\infty (S e^{\mu+\sqrt{\bar{m}^2T}a}-K) \frac{d^4}{da^4}e^{-a^2} da,
\end{eqnarray*}
which yields
\begin{eqnarray}
{\cal C}_2(S,T)=
SN(d_1)+\frac{Ke^{-rT}}{\sqrt{m^2T}}N'(d_2)\left[\frac{1}{2\bar{m}^2T}H_2\left(d_2/\sqrt{2}\right)-\frac{1}{\sqrt{2\bar{m}^2T}}H_1\left(d_2/\sqrt{2}\right)+1\right].
\label{c32}
\end{eqnarray}
We sum up all contributions given by Eqs.~(\ref{c12})--(\ref{c32}) and plugging them into Eq.~(\ref{c}) we finally obtain Eq.~(\ref{ct}) of the main text.


\begin{thebibliography}{000}
\bibitem{hull}J.C. Hull, Options, Futures, and other derivatives, Prentice Hall, New York, 1997.
\bibitem{osborne} M. F. M. Osborne, Operations Research {\bf 7} (1959) 145-173.
\bibitem{black.scholes} F. Black and M. Scholes, The Journal of Political Economy {\bf 81} (1973) 637--654.
\bibitem{hull87} J. C. Hull and A. White, {\it The pricing of options on assets with stochastic volatilities}, The Journal of Finance {\bf 42} (1987) 281--300.
\bibitem{Stein} E. Stein and J. Stein, {\it Stock Price Distributions with Stochastic Volatility: An Analytic Approach}, Review of Financial Studies {\bf 4} (1991) 727--752; reprinted in {\it Volatility: New Estimation Techniques for Pricing Derivatives}, Robert Jarrow Ed., Risk Publications, 1998, pp. 325--339.
\bibitem{Heston} S. Heston, {\it A closed-form solution for options with stochastic volatiliy with applications to bond and currency options}, Review of Financial Studies {\bf 6} (1993) 327--343.
\bibitem{scott} L. Scott, {\it Option Pricing when the Variance changes randomly: {T}heory, {E}stimation, and an {A}pplication}, J. Financial and Quantitative Analysis {\bf 22} (1987) 419--438.
\bibitem{fouque} J.-P. Fouque, G. Papanicolaou and K. R. Sircar {\it Mean-reveting stochastic volatility}, International Journal of Theoretical and Applied Finance {\bf 3} (2000) 101--142.
\bibitem{fouquebook} J.-P. Fouque, G. Papanicolaou, and K. R. Sircar, {\it Derivatives in Financial Markets with Stochastic Volatility} (Cambridge University Press, Cambridge, 2000).
\bibitem{ijtaf} J. Masoliver and J. Perell\'o, {\it A correlated stochastic volatility model measuring leverage and other stylized facts}, International Journal of Theoretical and Applied Finance {\bf 5} (2002) 541--562.
\bibitem{bouchaud} J. Perell\'o, J. Masoliver and J.-P. Bouchaud, {\it Multiple time scales in volatility and leverage correlations: a stochastic volatility model}, Applied Mathematical Finance {\bf 11} (2004) 27--50.
\bibitem{qf} Masoliver, J., J. Perell\'o, {\it Multiple time scales and the exponential Ornstein-Uhlenbeck stochastic volatility model}, Quantitative Finance 6 (2006) 423-433.
\bibitem{bacry} J.F. Muzy, J. Delour, E. Bacry, {\it Modelling fluctuations of financial time series: from cascade process to stochastic volatility model}, The European Physical Journal B {\bf 17} (2000) 537--548.
\bibitem{bouchaud-lev} J.-P. Bouchaud, A. Matacz, and M. Potters, {\it Leverage effect in financial markets: The retarded volatility model}, Physical Review Letters {\bf 87} (2001) 228701.
\bibitem{remer} R. Remer, and R. Mahnke {\it Application of Heston model and its solution to the German DAX data}, Physica A {\bf 344} (2004) 236-239.
\bibitem{qiu} T. Qiu, B. Zheng, F. Ren, and S. Trimper, {\it Return-volatility correlation in financial dynamics}, Physical Review E {\bf 73} (2006) 065103 .
\bibitem{spagnolo} G. Bonanno, D. Valenti, and B. Spagnolo, {\it Mean escape time in a system with stochastic volatility}, Physical Review E {\bf 75} (2007) 016106
\bibitem{buchbinder} G.L. Buchbinder, K.M. Chistilin, {\it Multiple time scales and the empirical models for stochastic volatility}, Physica A {\bf 379} (2007) 168--178.
\bibitem{biro} T.S. Bir\'o, R. Rosenfeld, {\it Microscopic origin of non-Gaussian distributions of financial returns}, Physica A {\bf 387} (2008) 1603-1612.
\bibitem{montagna} E. Cisana, L. Fermi, G. Montagna, O. Nicrosini, {\it A Comparative Study of Stochastic Volatility Models}, arXiv:0709.0810v1.
\bibitem{pre-lev} J. Perell\'o and J. Masoliver, {\it Random diffusion and leverage effect in financial markets}, Physical Review E {\bf 67} (2003) 037102.
\bibitem{comp} J. Perell\'o, J. Masoliver and N. Anento, {\it A comparison between correlated stochastic volatility models}, Physica A {\bf 344} (2004) 134--137.
\bibitem{eisler} Z. Eisler, J. Perell\'o, J. Masoliver, {\it Volatility: a hidden Markov process in financial time series}, Physical Review E {\bf 76} (2007) 056105 (11 pages).
\bibitem{extreme} J. Masoliver, J. Perell\'o, {\it Extreme times for volatility process}, Physical Review E {\bf 75} (2007) 046110 (11 pages).
\bibitem{pere2007} J. Perell\'o, {\it Market memory and fat tail consequences in option pricing on the expOU stochastic volatility model}, Physica A {\bf 382} (2007) 213-218.
\bibitem{ben} D. Ben-Avraham and S. Havlin, {\it Diffusion and Reactions in Fractals and Disordered Systems} (Cambridge University Press, Cambridge, 2000).
\bibitem{bouchaudbook} J.-P. Bouchaud and M. Potters {\it Theory of Financial Risk and Derivative Pricing: From Statistical Physics to Risk Management} (Cambridge University Press, Cambridge, 2003).
\bibitem{lo} A. Lo, {\it Long term memory in stock market prices}, Econometrica {\bf 59} (1991) 1279--1313.
\bibitem{ding} Z. Ding, C. W. J. Granger, and R. F. Engle, {\it A long memory property of stock market returns and a new model}, Journal of Empirical Finance {\bf 1} (1993) 83--106.
\bibitem{stanley} Y. Liu, P. Gopikrishnan, P. Cizeau, C.K. Peng, M. Meyer and H.E. Stanley, {\it Statistical properties of the volatility of price fluctuations}, Physical Review E {\bf 60} (1999) 1390--1400.
\bibitem{black76} F. Black, {\it Studies of stock market changes}, Proceedings 1976 of the American Statistical Association, Business and Economics Statistics (1976) Section 177.
\bibitem{bollerslev} T. Bollerslev, J. Litvinova, G. Tauchen, Leverage and Volatility Feedback Effects in High-Frequency Data, {\it Journal of Financial Econometrics} {\bf 4} (2006) 353--384.
\bibitem{barndorff} O. E. Barndorff-Nielsen, N. Shephard, {\it Non-Gaussian Ornstein–Uhlenbeck-based models and some of their uses in financial economics}, Journal of the Royal Statistical Society: Series B (Statistical Methodology) {\bf 63} (2001) 167--241.
\bibitem{lebaron} B. LeBaron, {\it Stochastic volatility as a simple generator of apparent financial power laws and long memory}, Quantitative Finance {\bf 1} (2001) 621–-631.
\bibitem{risken} H. Risken, {\it The Fokker-Planck equation} (Springer, Berlin, 1984).
\bibitem{Pochart} B. Pochart and J.-P. Bouchaud, {\it The skewed multifractal random walk with applications to option smiles}, Quantitative Finance {\bf 2} (2002) 303--314.
\bibitem{nonideal} J. Perell\'o and J. Masoliver, {\it The effect of non-ideal market conditions on option pricing}, Physica A {\bf 308} (2002) 420-442.
\bibitem{correlated} J. Masoliver and J. Perell\'o, {\it Option pricing and perfect hedging on correlated stocks}, Physica A {\bf 330} (2003) 622-652.
\end{thebibliography}
\end{document}